# Antisite defect qubits in monolayer transition metal dichalcogenides

Jeng-Yuan Tsai[1], Jinbo Pan[1], Hsin Lin[2], Arun Bansil[3]*, Qimin Yan[1]*

[1]Department of Physics, Temple University, Philadelphia, PA 19122, USA

[2]Institute of Physics, Academia Sinica, Taipei, Taiwan

[3]Physics Department, Northeastern University, Boston, MA 02115, USA

* Correspondence and requests for materials should be addressed to Q. Y. (qiminyan@temple.edu) and A.B. (ar.bansil@northeastern.edu)

**Being atomically thin and amenable to external controls, two-dimensional (2D) materials offer a new paradigm for the realization of patterned qubit fabrication and operation at room temperature for quantum information sciences applications. Here we show that the antisite defect in 2D transition metal dichalcogenides (TMDs) can provide a controllable solid-state spin qubit system. Using high-throughput atomistic simulations, we identify several neutral antisite defects in TMDs that lie deep in the bulk band gap and host a paramagnetic triplet ground state. Our in-depth analysis reveals the presence of optical transitions and triplet-singlet intersystem crossing processes for fingerprinting these defect qubits. As an illustrative example, we discuss the initialization and readout principles of an antisite qubit in $WS_2$, which is expected to be stable against interlayer interactions in a multilayer structure for qubit isolation and protection in future qubit-based devices. Our study opens a new pathway for creating scalable, room-temperature spin qubits in 2D TMDs.**

The ongoing second quantum revolution calls for exploiting the laws of quantum mechanics in transformative new technologies for computation and quantum information sciences (QIS) applications.[1] Spin-qubits based on solid-state defects have emerged as promising candidates because these qubits can be initialized, selectively controlled, and readout with high fidelity at ambient temperatures.[2,3] Solid-state defects, especially in 2D TMDs offer advantages of scalability and ease of device fabrication. Point defects as spin qubits have been demonstrated in traditional semiconductor systems,[4] including the nitrogen-vacancy ($NV^-$) center in diamond and the spin-1/2 defect in doped silicon,[3,5,6,7,8,] among other possibilities[4,9-11] In particular, Si-vacancy complex in diamond,[11] vacancy defects in SiC,[12] and vacancy complexes in $AlN^{10}$ have been predicted as qubits. A neutral divacancy $(V_C-V_{Si})^0$ in SiC has been identified as a qubit with millisecond coherence time,[13] where an improvement in dephasing time by over four orders of magnitude can be achieved by embedding the qubit in a decoherence-protected subspace through microwave dressing.[14]

A key challenge in the development of controllable multiple-qubit systems is how to effectively couple spin defects and achieve high fidelity and long coherence times. The planar structures of atomically-thin 2D materials present a superior platform for realizing controlled creation and manipulation of defect qubits with better potential for scalability than the bulk materials. In 2D materials, defects can be generated by a number of existing approaches,[15] and characterized and manipulated using atomic-level scanning probe techniques.[16] The carbon-vacancy complex ($C_B$-$V_N$) in hexagonal boron nitride (h-BN) has emerged as the first such qubit.[17] Nitrogen vacancy complex defect ($N_B$-$V_N$) and the negatively charged boron vacancy defect ($V_B$) have also been proposed as qubit candidates in h-BN,[18, 19] and a number of point defects in h-BN show promise as ultra-bright single-photon emitters at room temperature.[20,21]

TMDs are a major class of 2D graphene cognates that are attracting intense current interest because of their sizable band gaps and high absorption coefficients, among other unique physical and chemical properties. Atomic defects in as-grown TMD samples, such as the anion vacancies,[22] are well known to play an essential role in their electronic behavior.[23] Compared to 3D wide-band-gap materials, the spin coherence time in $MoS_2$ has been estimated to be extremely long, on the order of 30 ms, suggesting the potential of TMDs as good host materials for multiple-qubit operation.[24] The field is thus ripe for the discovery and rational design of novel defect qubits in 2D materials and their implementation in single- and multi-qubit platforms for QIS applications.

Here, we report the identification of anion antisite defects $M_X$ in six $MX_2$ (M = Mo, W; X = S, Se, Te) TMD systems as novel 2D solid-state defect qubits obtained via a high-throughput search based on a new qubit formation hypothesis involving symmetry constraints as well as the host electronic structures. Our first-principles defect computations, see the Methods section below for details, demonstrate that the proposed antisites in these TMDs host paramagnetic triplet ground states with flexible level splittings controlled by site symmetries and both the in-plane and out-of-plane $d$ orbital interactions. Taking $W_S$ antisite in $WS_2$ as an especially favorable case, we demonstrate a viable transition loop between the triplet and singlet defect states, including optical excitations/relaxations and nonradiative decay paths for the $W_S$ antisite as a qubit. A complete set of qubit operational processes, including initialization, manipulation, and readout steps is delineated to provide a blueprint for experimental verification.

**Qubit discovery hypothesis**

Our data-driven defect qubit discovery effort in the TMDs is based on satisfying three major descriptors as follows. (1) A paramagnetic, long-lived triplet ground state with multiple in-gap defect levels. (2) An optical transition path between the ground and excited triplet states as well as a spin-selective (nonradiative) decay path between the different spin-multiplet states for qubit initialization. And, (3) distinguishable luminescence signatures for the two spin sublevels for qubit readout.[4]

Before we turn to discuss how the interplay of the host electronic structure and local site symmetry yields an anion antisite in the TMDs as a viable defect qubit, we note that wide bandgap compounds, such as SiC, AlN, and h-BN, are mostly characterized by occupied anion states as valence bands and unoccupied cation states as conduction bands. As a result, cation (anion) vacancy defect levels that originate from anion (cation) dangling-bond states are usually located in the valence (conduction) band. Therefore, it becomes necessary to introduce impurities next to the vacancies[4] or apply strain perturbations[10] in the wide bandgap systems to create additional energy splittings to push the defect levels into the gap. Monolayer group-VI TMDs possess fundamentally different electronic structures that are characterized by dominant $d$-states of the transition metals with $d_{z^2}$ and $\{d_{xy}, d_{x^2-y^2}\}$ states contributing to the valence and conduction band edges, respectively.[25] As a result, point defects created by cations such as the anion antisites and anion vacancies/complexes are more likely to host deep in-gap defect levels. Notably, intrinsic defects including vacancies and anion antisites have been observed experimentally in the TMDs.[26]

It is useful to recall here that a triplet ground state is preferred (Hund's rule) when the exchange energy involving the interaction of two parallel spins is favorable compared to the energy required to lift one of the electrons to a higher level. In other words, either a small or zero energy splitting between the two highest occupied levels is a prerequisite for stabilizing the triplet ground state. An energetically favorable scenario is that the local site-symmetry of the point defect belongs to a point group with at least one 2-dimensional (2D) irreducible representation (irrep). The 2D irrep may generate doubly degenerate defect levels and hence a strong tendency to create a triplet ground state when these two levels are the highest occupied levels as is the case for the NV center in diamond. $d$ states of transition metal cations in TMDs tend to have relatively large exchange energies, which favors a triplet ground state in keeping with the Hund's rule.

**Antisites in group-VI TMDs: a novel qubit platform**

Based on the preceding discussion of our hypothesis, we performed a symmetry-based data-mining search to identify nonmagnetic and relatively stable MX$_2$ TMD compounds from the 2D materials database C2DB[27] with computed band gaps larger than 1.4 eV and energies above the convex hull less than 0.1 eV/atom. 27 TMD compounds were identified and assigned to three different phases, 1H, 1T, and 1T' and the corresponding point groups, D$_{3h}$, D$_{3d}$, and C$_{2h}$, respectively. 1T' phase is ruled out since the C$_{2h}$ point group does not have a 2D irrep. For group-VI TMDs, the 1H phase is more stable than the 1T phase under equilibrium conditions.[28] Therefore, we focused on six nonmagnetic group-VI 1H TMDs MX$_2$ (M: Mo, W; X: S, Se, Te). We performed high-throughput defect computations and found that no anion vacancy in these six group-VI TMDs hosts a triplet ground state, which is largely due to the fact that the cation dangling bond states surrounding anion vacancies form a state space consisting of fully occupied 1D $a_1$ and unoccupied 2D e levels. As a result, the anion vacancies do not favor a triplet spin configuration. We thus ruled out isolated anion vacancies and focused on anion antisite defects in the 1H TMDs.

Figure 1(a) presents an example of an anion antisite in TMDs where a cation is located on an anion site in the crystal lattice. The location and occupation of defect levels created by the antisite are controlled by its interaction with the three cation atoms in the central atomic layer as well as the defect charge state (see Methods). Defect levels of six anion antisites $M_X^0$ in the band gaps of 1H-MX$_2$ (M: Mo, W; X: S, Se, Te) were computed and those of the neutral antisites are shown in Figure 1(b). It is remarkable that all six 1H TMDs host neutral antisites in a triplet ground state. Note that it has been predicted that anion antisite in WS$_2$[29] and MoS$_2$[30] favors a triplet state. In spite of the universal presence of a triplet ground state in these six antisite systems, we emphasize that the level splittings for the three defect levels in the spin-up channel are not universal (Figure 1(b)). Constructed mainly from $d_{x^2-y^2}$, $d_{xy}$, and $d_{z^2}$ orbitals of the cation atoms located at the antisites, the three defect levels of the six 1H TMDs fall into two level-splitting patterns characterized by the position of the $d_{z^2}$ level relative to the $d_{x^2-y^2}$ and $d_{xy}$ levels. For the neutral antisites in MoS$_2$, MoSe$_2$, WS$_2$, and WSe$_2$, the two highest occupied levels in the gap are doubly degenerate, while those in MoTe$_2$ and WTe$_2$ generate three discrete defect levels in the band gaps of the host material.

In order to gain insight into the differences in defect-level splittings in various 1H TMDs, we adopt a local-symmetry analysis. The local symmetry for the unperturbed environment of an antisite is $C_{3v}$, which is a subgroup of the crystal point-group $D_{3h}$ of the pristine 1H-$MX_2$ systems. In 1H-$MX_2$ compounds, the $d$ orbitals hybridize and transform into 2D irrep E and 1D irrep $A_1$. Within this local symmetry, $d_{x^2-y^2}$ and $d_{xy}$ orbitals at the antisites belong to the 2D irrep E while the $d_{z^2}$ orbital belongs to the 1D irrep $A_1$. The doubly degenerate $d_{x^2-y^2}$ and $d_{xy}$ orbitals at the ansitites interact with the three neighboring cations mainly in the $x$-$y$ plane. Their energy levels are therefore affected by the M-M distances. On the other hand, the interaction between the $d_{z^2}$ orbitals of the antisites and the three cations is determined by the location of the antisite defect along the $z$-direction relative to the cation layer.

Our analysis indicates that, due to differences in orbital interactions, the $d_{z^2}$ defect level shifts up in energy relative to the $d_{x^2-y^2}$ and $d_{xy}$ levels as the antisite moves away from the cation layer along the z-direction. This relative shift in energies eventually reaches a critical point where the level switching takes place. However, as shown in Figure 1(c), the position of the antisite along the $z$-direction in equilibrium structure (red lines) is negatively correlated with the lattice constant. Due to differences in the critical $z$-positions (yellow lines), where the levels undergo switching, two level-splitting patterns emerge depending on the lattice constants of the host materials. Since the lattice constants of the six 1H TMDs are mainly determined by anion species, we see a clear correlation between the level-splitting pattern and anion species. For antisites in $MoS_2$, $MoSe_2$, $WS_2$, and $WSe_2$, the $d_{x^2-y^2}$ and $d_{xy}$ levels are located below the $d_{z^2}$ level (2/1 splitting), while the opposite is the case for $MoTe_2$ and $WTe_2$ (1/2 splitting).

Due to interactions between the antisite and adjacent cation atoms, the neutral antisite has two extra electrons after bonding which can occupy two defect levels in the gap, creating two occupied levels and one unoccupied level in the up-spin channel in the triplet state. In the case of 2/1 splitting, the lowest two levels $d_{x^2-y^2}$ and $d_{xy}$ are occupied and remain doubly degenerate. In the case of 1/2 splitting ($MoTe_2$ and $WTe_2$), in contrast, a single electron occupying $d_{x^2-y^2}$ and $d_{xy}$ levels introduces a sizable spontaneous Jahn-Teller distortion which reduces the site-symmetry - $C_{3v}$ approximately to the point-group symmetry $C_h$ (see Table S1 in SI) and pushes the occupied $d_{x^2-y^2}$ level down and closer to the $d_{z^2}$ level. We emphasize that different level-splitting patterns of neutral antisites in the 1H TMD family originate from the unique anisotropic orbital interactions in 2D materials.

In order to evaluate the stabilities of neutral antisite defects in the six 1H TMDs, we calculate the thermodynamic charge transition levels shown in Figure 1(d). Charge state corrections for the charged antisite systems are adopted by utilizing an extrapolation method (see Methods).[31,32] Energy windows for the Fermi level in the gap where the neutral charge state is most stable are 1.43 eV, 0.88 eV, 0.48 eV, 1.01 eV, 1.24 eV, 0.19eV for $W_S^0$, $W_{Se}^0$, $W_{Te}^0$, $Mo_S^0$, $Mo_{Se}^0$, and $Mo_{Te}^0$, respectively (Figure S1 in SI). The thermodynamic transition levels, obtained via the hybrid functional calculations, are expected only to capture the trends in defect stabilities. However, it is reasonable to expect that neutral antisite defects will be stable in these 1H-TMDs when the Fermi level is close to the mid-gap.

For a viable defect qubit, the defect levels related to qubit operation must be deep in the gap to minimize effects of disruptive interactions with the bulk bands. The highest occupied defect levels of $Mo_S^0$, $Mo_{Se}^0$, $Mo_{Te}^0$, and $W_{Te}^0$ lie close (within 0.3 eV) to the valence band maximum (VBM) of the host materials. In contrast, the defect levels in $W_S^0$ and $W_{Se}^0$ are sufficiently deep (about 0.6 eV above the VBM) for qubit operation.[33] Among the six anion antisites, $W_S^0$ and $W_{Se}^0$ are therefore the most promising candidates as novel defect qubits in 2D TMDs.

**Antisite defect qubit in WS$_2$**

We now discuss $W_S^0$ in WS$_2$ as a benchmark system to demonstrate the operation principle of antisite defect qubits. A pristine monolayer of WS$_2$ in the H-phase (Table S1 in the SI) is composed of three hexagonal layers that form a sandwich-like structure (S-W-S). We define the direction of the *c* lattice vector as the *z*-axis. The sulfur atoms occupy the upper and lower hexagonal sublattice sheets with a symmetrical W plane lying between these sheets. Our optimized structure indicates that the W-S and S-S distances are 2.391 Å and 3.107 Å, respectively.[28] Hybrid functional calculations predict a band gap of 2.40 eV which is close to the experimental value of about 2.41 eV.[34] The optimized structure of the anion antisite $W_S^0$ is shown in Figure 2(a). The local environment of $W_S^0$ without perturbation has $C_{3v}$ symmetry with the rotation axis lying along the z-direction. Note that if the antisite in WS$_2$ is initially perturbed by a random displacement, the symmetry of the resulting structure can be lowered from $C_{3V}$ to $C_h$ with lower energy by ~25 meV per unit cell compared to the metastable structure with $C_{3V}$ symmetry, see Section 4 in SI for details.

The calculated electronic structure of $W_S^0$ hosts a triplet ground state. The in-gap defect levels can be labeled by irreps of the point-group $C_{3v}$ as shown in Figure 2(b). The ground and excited states of $W_S^0$ are described by single Slater determinants as $e^2$ and $a_1^1 e^1$, respectively. It is noted that one can equivalently express a many-body state by either electron occupation or hole occupation of the single-particle orbitals.[35] From this point of view, the defect levels of NV$^-$ center in diamond (hole occupation) are identical to the defect levels of $W_S^0$ (electron occupation) in terms of single Slater determinants.[35,36] Therefore, we will adopt the state symbols for $e^2$ and $a_1^1 e^1$ as { $^3A_2$, $^1E$, $^1A_1$} and $^3E$, respectively.

In order to access transition processes involving the triplet ground state $^3A_2$ and the triplet excited state $^3E$, we performed constrained DFT (CDFT) calculations in which occupations of the Kohn-Sham orbitals are constrained to desirable configurations. We set the occupation of defect levels as $e^{0.5} e^{0.5} a_1^1$ for the excited state to circumvent the Jahn-Teller distortion and preserve the $C_{3v}$ symmetry.[37] This methodology combined with HSE calculations has been applied to evaluate the zero-phonon line (ZPL) of NV$^-$ center to achieve excellent agreement with experiments.[4] As shown in the configuration coordinate diagram (Figure 2(c)), the ZPL for the internal transition between the triplet ground and excited states is 0.729 eV, which lies in the near-infrared (IR) range. Frank-Condon relaxation energies are 5 meV and 0.8 meV for the excited and ground states, respectively. The Huang-Rhys factors are computed to be 0.0004 and 0.0013 for ground and excited state, respectively. The inclusion of spin-orbit coupling shifts the ZPL to a slightly lower energy but only induces minor changes in the relaxation energies and Huang-Rhys factors (see Section 5 in SI for details). These extremely small vibrational couplings in the internal transitions imply that antisite defects in TMDs will likely be suitable for other QIS applications such as single-photon emitters and quantum sensors.

Positions of singlet states are significant for nonradiative decay paths that connect triplet and singlet states. We estimate positions of singlet states $^1E$ and $^1A_1$ by considering the Coulomb interaction.[36,38] Note that since the singlet state $^1A_1$ is a strongly correlated state, it cannot be described accurately as a single-particle Kohn-Sham state. Since we have the same local symmetry, we can adopt the results for the NV⁻ center[36] based on group theory in which the ratio of energy shifts for $^1E$ and $^1A_1$ relative to $^3A_2$ is 1:2. The energy difference obtained by first-principles calculations between $^1E$ and $^3A_2$ is 0.275 eV, from which the energy difference between $^1A_1$ and $^3A_2$ can be estimated to be 0.55 eV. Considering that the ZPL of the triplet states is 0.729 eV, it indicates that singlet states $^1E$ and $^1A_1$ are located between the triplet excited state $^3E$ and the triplet ground state $^3A_2$ (Figure 2(d)).

In order to operate as a qubit, a defect center must have distinct signatures of optical transitions involving various sublevels and support nonradiative decay paths.[12] The spin-orbit coupling (SOC) effects and the associated sublevels are important for ascertaining allowed intersystem crossings (ISCs) between different spin configurations.[39] We analyzed the matrix element of single-particle spin-orbit operator $H_{so}$ with the framework of group theory to determine the symmetry-allowed intersystem crossings, see Section 3 in SI for a detailed analysis. Three allowed intersystem crossing paths $\Gamma_0^\perp$, $\Gamma_1^\perp$, and $\Gamma_2^z$ are identified based on spin quantum numbers and irreps of tensor products of wavefunctions and spinors. The allowed spin-conserving optical transitions and intersystem crossings are shown in Figure 2(d). We denote $\Gamma_0^\perp$ and $\Gamma_1^\perp$ involving the nonaxial components of $H_{so}$, while $\Gamma_2^z$ involves the axial component.

**Qubit operation principle**

A complete loop for qubit operation based on $W_S^0$ in $WS_2$ is illustrated in Figure 3. The initialization, manipulation, and readout of our TMD-based antisite qubit resemble the operations of defect qubits in the well-known NV⁻ center in diamond.[4,12,40] We choose the sublevel $A_1$ ($m_s = 0$) and one of sublevels in E ($m_s = \pm 1$) in the triplet ground state $^3A_2$ as a two-level qubit system. Initialization of the qubit could then be achieved by optically pumping the defect center from the sublevel E in the triplet ground state $^3A_2$ to the sublevel $A_1$ in triplet excited state $^3E$. The sublevel $A_1$ in the triplet excited state has an allowed intersystem crossing to the sublevel $A_1$ in the singlet state $^1A_1$ via path $\Gamma_0^\perp$, and then the system relaxes back to the sublevel $A_1$ in the triplet ground state via path $\Gamma_2^z$. The preceding transition processes form a complete cycle for the initialization of the qubit.

Manipulation of the qubit could be implemented by utilizing one of the sublevels E and the sublevel $A_1$ in the triplet ground state by applying resonant microwave. The fluorescence intensity from sublevels E is expected to be weaker than that from the sublevel $A_1$ due to the existence of the intersystem crossing path $\Gamma_0^\perp$. Readout of the qubit can therefore be realized by detecting the difference in fluorescence intensity involving different qubit states. The set of operations presented above would enable qubit initialization, manipulation, and readout, forming the essential operation principles for antisite qubits in TMDs.

**Qubit protection scheme and spin coherence**

Our antisite qubits, which involve TMD monolayers will be susceptible to the presence of molecules, ions, and other chemical species in the environment. In order to address this problem, we have investigated a qubit protection scheme (Figure 4(a)) in which the $MX_2$ monolayer is capped on both sides with a layer of hexagonal boron nitride (h-BN) as a protective cover. Based on our first-principles computations using the hybrid functional with the standard mixing parameter and the inclusion of van der Waals corrections (optPBE-vdW) for structural relaxation,[41,42] we find that the triplet ground state is preserved for the antisite qubit in the h-BN/$WS_2$/h-BN heterojunction (Figure 4(b)). Energy separation between the highest occupied and the lowest unoccupied defect level in the spin-up channel (related to the ZPL energy) is around 1.1 eV, which is close to that in the monolayer system. A small level splitting of 0.045 eV is observed between the two occupied defect levels which is associated with the slight symmetry breaking induced by the neighboring h-BN layer.

The preceding observations indicate the effectiveness of adopting h-BN as protection layers to isolate the antisite qubits in monolayer TMDs from both environmental and substrate effects. As shown in Figure 4(c), the projected density of states (PDOS) on h-BN layers is located deep in the conduction and valence bands of the heterojunction due to the large band gap of h-BN (~ 6 eV) and the type-I band alignment between h-BN and WS$_2$.[43] We thus expect that the key optical transitions related to the qubit operation will not be significantly affected by the h-BN isolation/protection layers. The protected antisite qubits in 2D heterostructures thus offer a novel and robust platform for quantum information technologies.

Another key factor concerns the spin decoherence time T$_2$ of a qubit. Taking MoS$_2$ as an example, the previous work[24] has shown that the decoherence of the electron spin originates mainly from the presence of $^{95}$Mo and $^{97}$Mo cation nuclear spins of 5/2, and that it can be greatly diminished by utilizing nuclear-spin-free isotope for which an exceptionally long spin decoherence time (more than 30 ms) has been predicted, owing to molybdenum's small gyromagnetic ratio. The gyromagnetic ratio $\gamma(^{183}W)/\gamma(^{95}Mo)$ is 0.64, and therefore, even longer spin decoherence time can be expected in WS$_2$ and WSe$_2$ based defect qubits for realizing controllable multi-qubit operations in solid-state 2D systems.

**Summary and discussion**

Using a high-throughput materials discovery effort based on a defect-qubit design hypothesis involving the interplay of local symmetry of the defect and the electronic structure of the host, we identify thermodynamically stable, neutral anion-antisite defects in six monolayer 1H-MX$_2$ TMD compounds as potential defect-spin qubits hosting stable triplet ground states. The optical signatures of these qubits, including the ZPL for optical transitions, are evaluated using an in-depth analysis of the electronic configurations and the corresponding symmetry representations of the defect states in the antisites. Intersystem crossing channels for qubit initialization and operation are identified. A scheme for isolating and protecting the antisite qubits is proposed based on a h-BN/TMD/h-BN heterojunction structure. Our study opens a new pathway for creating spin-qubits and multi-qubit platforms for quantum information technologies based on defects in 2D solid-state systems.

**Methods**

*Computational details.* All calculations were performed by using the Vienna Ab initio Simulation Package (VASP)[44] based on the density functional theory (DFT)[45,46]. To calculate the spin density near the nuclei, the projector-augmented-wave method (PAW)[47,48] and a plane-wave basis set were used. Recent advances using hybrid functionals have led to accurate descriptions of defect states by overcoming the well-known band-gap problem of the traditional DFT. Our calculations were performed using the screened hybrid-functional of Heyd-Scuseria-Ernzerhof (HSE)[49,50] with default mixing parameter and the standard range-separation parameter (0.2 Å$^{-1}$) to reproduce the experimental quasiparticle gap of pristine WS$_2$.[34] The plane-wave basis set energy cutoff was set to 320 eV. For defect supercell calculations, we adopt Γ point in the Brillouin zone for defect-state calculations to avoid undesirable splitting of defect states. For charged defect formation energy calculations, a special k-point at (0.25, 0.25, 0) in the first Brillouin zone is adopted (see Section 2 in SI for details). A vacuum space of 20 Å is added along the direction perpendicular to the monolayer, and the planar size of the supercell is 5 × 5 in order to avoid interactions between adjacent images. Structural relaxations have been performed for all the systems investigated which were converged until the force acting on each ion was less than 0.01 eV/Å. The convergence criteria for total energies for structural relaxations and self-consistent calculations are $10^{-4}$ eV and $10^{-5}$ eV, respectively. The constrained DFT (CDFT) methodology[51,52] was employed for the calculation of excitation energies between triplet states.

*Defect formation and transition levels.* Relative stability of point defects depends on the charge states of the defects. We analyze this issue for antisite defects in TMDs by calculating the defect formation energy ($E_f$) for charge state q, which is defined as: $E_f(\epsilon_F) = E_{tot}^q - E_{bulk} + \mu_X - \mu_M + q(\epsilon_F + E_v) + \Delta E$, where $E_{tot}^q$ is the total energy of the charged defect system with charge q, $E_{bulk}$ is the total energy of the perfect MX$_2$ system, $\mu_M$ is the chemical potential of the metal atom M, $\mu_X$ is the chemical potential of the anion atom X, $\epsilon_F$ is the position of the Fermi-level with respect to the valence band maximum $E_v$, and $\Delta E$ is the charge correction energy. Transition levels are defined as $\epsilon(q'/q) = (E_f^{q'} - E_f^q)/(q - q')$, where $E_f^q$ is the formation energy for the state of charge q. We can interpret the transition levels as the Fermi level positions at which the formation energies of the defect in two distinct charge states are equal. The ionized energy of donor/acceptor is defined as the energy difference of transition level $\epsilon(+/0)/\epsilon(0/-)$ and CBM/VBM. In a low-dimensional system, due to anisotropic screening, ionization energy (IE) diverges with respect to the vacuum, and we adopted a charge correction method.[31,32] We assume that the chemical potential of M and X are in thermal equilibrium with MX$_2$, i.e., $\mu_{MX_2} = \mu_M + 2\mu_X$, where $\mu_{MX_2}$ is the energy of the perfect

MX$_2$ system. The accessible range of μ$_M$ and μ$_X$ can be further limited by the lowest energy phases of these elements depending on growth conditions. It should be noted that the transition levels do not depend on the choice of chemical potentials.

**Acknowledgments**

This work was supported by the U.S. Department of Energy, Office of Science, Basic Energy Sciences, under Award #DE-SC0019275. It benefitted from the supercomputing resources of the National Energy Research Scientific Computing Center (NERSC), a U.S. Department of Energy Office of Science User Facility operated under Contract No. DE-AC02-05CH11231.14

**Figure 1 Anion antisite defects in six 1H transition metal dichalcogenides.** (a) A schematic illustration of the $M_X^0$ antisite defect in monolayer 1H-TMD. (b) In-gap defect levels of $M_X^0$ in six 1H-TMDs with triplet ground states. Blue and orange colored bars represent the valence and conduction bands of the host materials. Note that $M_{S/Se}^0$ has doubly degenerate highest-occupied defect levels, whereas in the case of $M_{Te}^0$ there is a splitting between the occupied defect levels. (c) Correlation between the defect-level splittings and the $z$-positions of the antisites relative to those of the neighboring cations. Red lines mark the equilibrium $z$ positions of antisite defects, while the orange colored lines indicate the critical $z$ positions where a transition from the 2-1 type splitting (blue bars) to the 1-2 type splitting (light blue bars) takes place. (d) Thermodynamic transition levels for the six antisite defects in 1H-TMDs. $\epsilon(+/0)$ and $\epsilon(0/-)$ denote the transition levels from the charge state +1 to 0, and from 0 to -1, respectively. Neutral charge states are thermodynamically stable when the Fermi level is close to the mid-gap.

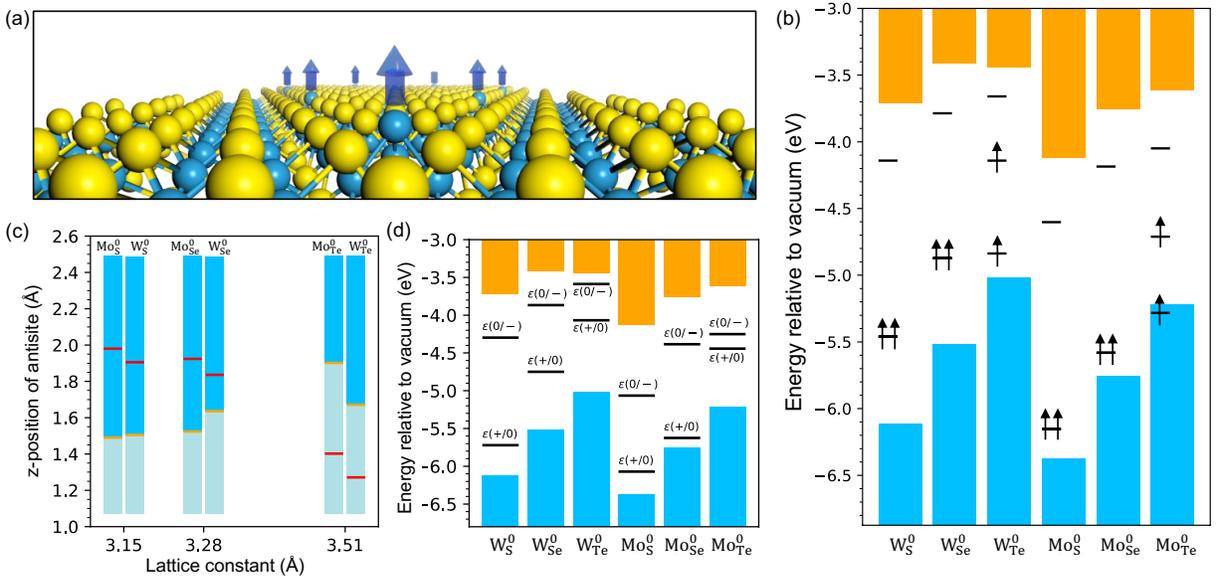

**Figure 2 Electronic and geometric structure of the neutral antisite defect $W_S^0$ in $WS_2$.** (a) Optimized structure of the antisite defect $W_S^0$ in $WS_2$, showing its $C_{3v}$ local symmetry. (b) Energy diagram showing the defect levels in the triplet ground state $^3A_2$. The defect levels e and $a_1$ are mainly composed of $\{d_{x^2-y^2}, d_{xy}\}$ and $d_{z^2}$ orbitals of the defect. The depictions of the wavefunctions involved are shown. (c) Configuration coordinate diagram of $W_S^0$ in $WS_2$ for the triplet ground state $^3A_2$ and the triplet excited state $^3E$. (d) Sublevels for the triplet ground state $^3A_2$, the triplet excited state $^3E$, and the singlet states $^1E$ and $^1A_1$, labeled by the irreps of $C_{3v}$. Spin-conserving optical transitions are shown by colored solid arrows. Symmetry-allowed intersystem-crossing paths are noted by dashed arrows. The labels $\{\Gamma_0^\perp, \Gamma_1^\perp\}$ and $\Gamma_2^z$ indicate the allowed intersystem-crossing paths via the nonaxial spin-orbit coupling and the axial spin-orbit coupling, respectively.

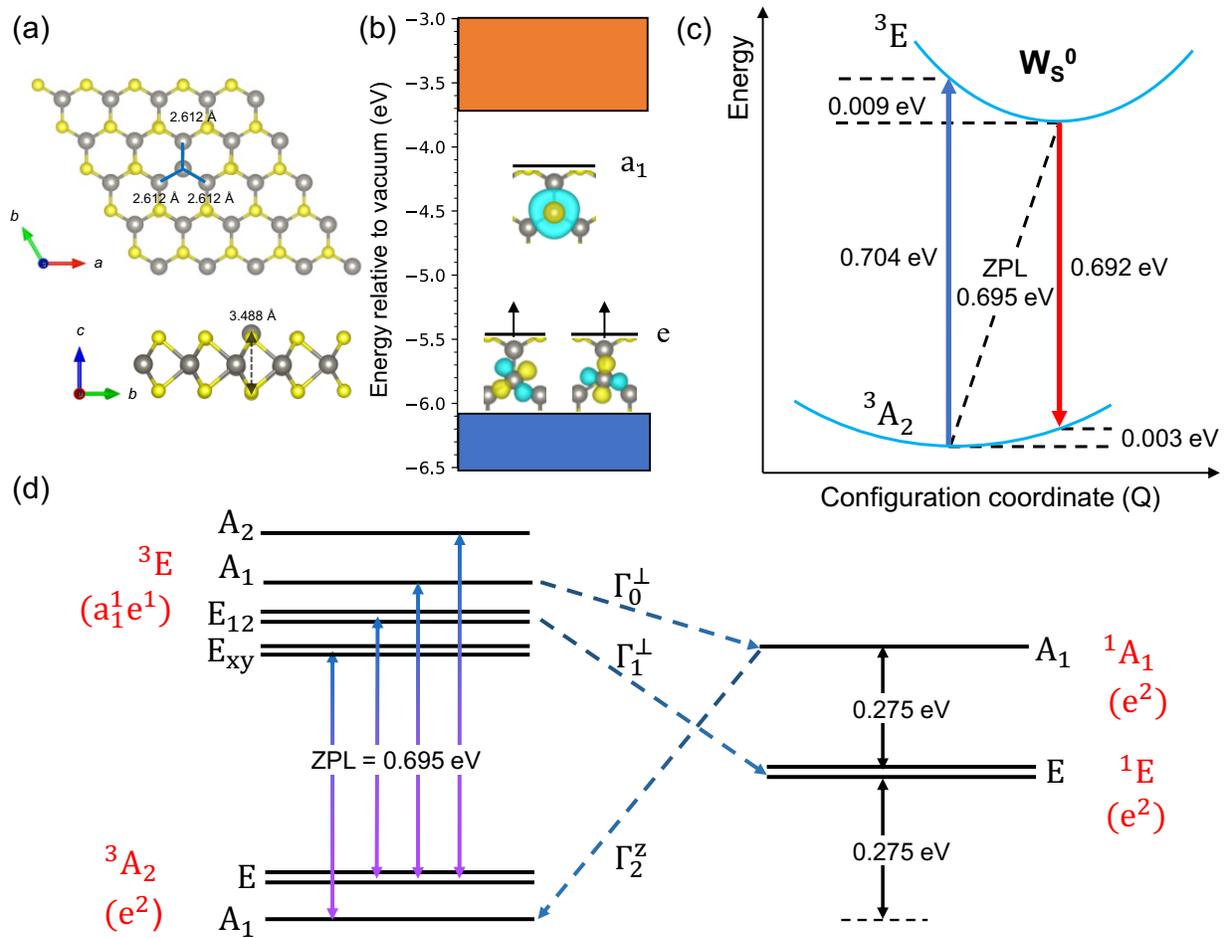

**Figure 3 Operational loop for the antisite qubit $W_S^0$.** Initialization (left panel): The defect center is pumped optically (solid red line) from the sublevel E in the triplet ground state $^3A_2$ to the sublevel $A_1$ in the triplet excited state $^3E$, and then the defect center relaxes back to the sublevel $A_1$ in the triplet ground state via the intersystem-crossing paths $\Gamma_0^\perp$ and $\Gamma_2^z$ (dashed purple lines). Manipulation (middle): The qubit can be manipulated by using electron paramagnetic resonance (EPR) on one of the sublevels E and the sublevel $A_1$ in the triplet ground state. The blue circular arrows indicate the manipulation process via a microwave pulse. Readout (right): The defect center is optically pumped again, and the intensities of fluorescence involving different initial states are detected. Note that the fluorescence process $E_{xy}(^3E) \rightarrow A_1(^3A_2)$ (thick orange line) has a higher intensity than the process $A_1(^3E) \rightarrow E(^3A_2)$ (thin orange line) due to the existence of intersystem crossing transitions (dashed purple lines) that weaken the radiative transition.

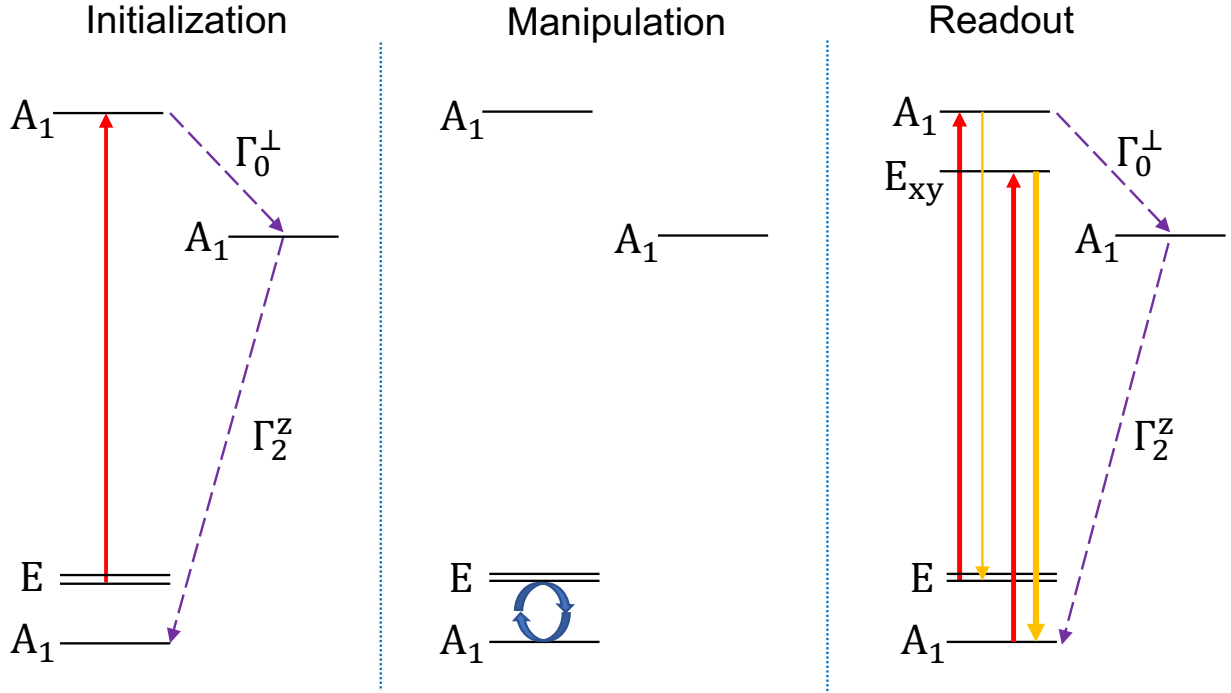

**Figure 4 Illustration of an environmentally protected qubit design based on the h-BN/WS$_2$/h-BN heterojunction structure.** (a) Schematic of the proposed 2D-heterojunction structure. (b) Schematic of the optimized heterojunction consisting of $2\sqrt{7} \times 2\sqrt{7}$ h-BN as the top and bottom layers and $4 \times 4$ WS$_2$ with antisite W$_S^0$ as the middle layer. The bottom h-BN layer isolates the qubit layer from the substrate, while the top h-BN layer provides protection against external environmental effects. (c) The in-gap defect levels where two electronic levels are occupied by spin-up electrons in a triplet ground state. (d) Computed density of states of the heterojunction and the projected density of states on B and N atoms indicating that the qubit can be optically initialized and readout without significant perturbation from the h-BN isolation/protection layers owing to the very large band gap of h-BN and the type-I band alignment of h-BN and WS$_2$.

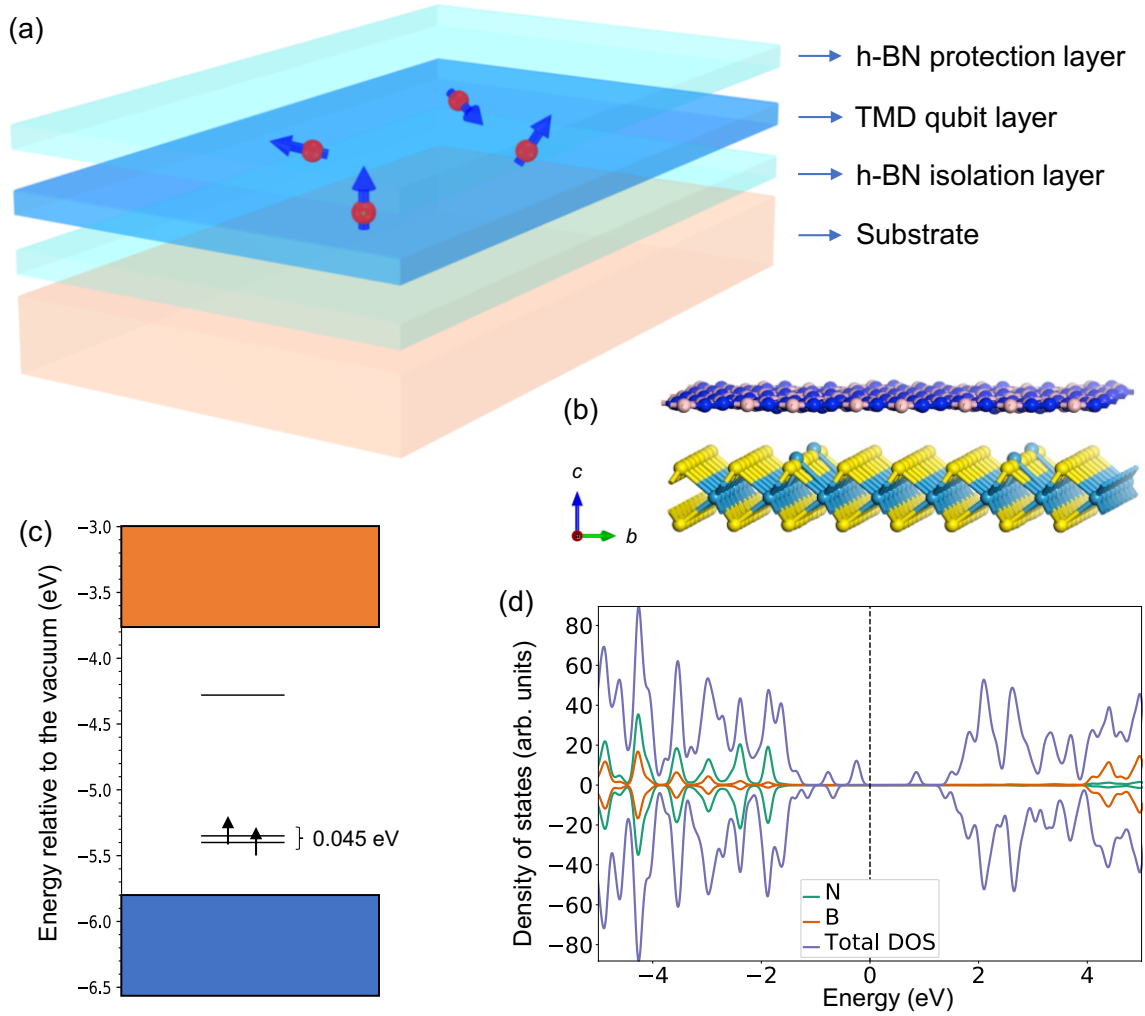

# Reference

1       Bharath, R. Quantum computing: a gentle introduction. *Choice: Current Reviews for Academic Libraries* **49**, 345-346 (2011).
2       Greentree, A. D., Fairchild, B. A., Hossain, F. M. & Prawer, S. Diamond integrated quantum photonics. *Materials Today* **11**, 22-31 (2008).
3       Pla, J. J. *et al.* A single-atom electron spin qubit in silicon. *Nature* **489**, 541-545, (2012).
4       Weber, J. R. *et al.* Quantum computing with defects. *Proceedings of the National Academy of Sciences,* **107**, 8513-8518 (2010).
5       Kane, B. E. A silicon-based nuclear spin quantum computer. *Nature* **393**, 133-137 (1998).
6       Salfi, J. *et al.* Quantum simulation of the Hubbard model with dopant atoms in silicon. *Nature Communications* **7**, 11342 (2016).
7       Doherty, M. W. *et al.* The nitrogen-vacancy colour centre in diamond. *Physics Reports* **528**, 1-45 (2013).
8       Jelezko, F., Gaebel, T., Popa, I., Gruber, A. & Wrachtrup, J. Observation of Coherent Oscillations in a Single Electron Spin. *Physical Review Letters* **92**, 076401 (2004).
9       Gali, A. Time-dependent density functional study on the excitation spectrum of point defects in semiconductors. *Phys. Status Solidi B* **248**, 1337-1346 (2011).
10      Seo, H., Govoni, M. & Galli, G. Design of defect spins in piezoelectric aluminum nitride for solid-state hybrid quantum technologies. *Scientific Reports* **6**, 20803 (2016).
11      Rogers, L. J. *et al.* All-Optical Initialization, Readout, and Coherent Preparation of Single Silicon-Vacancy Spins in Diamond. *Physical Review Letters* **113**, 263602 (2014).
12      Weber, J. R. *et al.* Defects in SiC for quantum computing. *Journal of Applied Physics* **109**, 102417 (2011).
13      Christle, D. J. *et al.* Isolated electron spins in silicon carbide with millisecond coherence times. *Nature Materials* **14**, 160-163 (2015).
14      Miao, K. C. *et al.* Universal coherence protection in a solid-state spin qubit. *Science* **369**, 1493 (2020).
15      Zhong, L. *et al.* Defect engineering of two-dimensional transition metal dichalcogenides. *2D Materials* **3**, 022002 (2016).
16      Wong, D. *et al.* Characterization and manipulation of individual defects in insulating hexagonal boron nitride using scanning tunnelling microscopy. *Nature Nanotechnology* **10**, 949 (2015).
17      Wu, F., Galatas, A., Sundararaman, R., Rocca, D. & Ping, Y. First-principles engineering of charged defects for two-dimensional quantum technologies. *Physical Review Materials* **1**, 071001 (2017).
18      Ivády, V. *et al.* Ab initio theory of the negatively charged boron vacancy qubit in hexagonal boron nitride. *npj Computational Materials* **6**, 41 (2020).
19      Vuong, T. Q. P. *et al.* Phonon-Photon Mapping in a Color Center in Hexagonal Boron Nitride. *Physical Review Letters* **117**, 097402 (2016).
20      Aharonovich, I. & Toth, M. Quantum emitters in two dimensions. *Science* **358**, 170 (2017).
21      Grosso, G. *et al.* Tunable and high-purity room temperature single-photon emission from atomic defects in hexagonal boron nitride. *Nature Communications* **8**, 705 (2017).

22   Tongay, S. *et al.* Defects activated photoluminescence in two-dimensional semiconductors: interplay between bound, charged, and free excitons. *Scientific Reports* **3**, 2657 (2013).
23   Wang, D., Li, X.-B., Han, D., Tian, W. Q. & Sun, H.-B. Engineering two-dimensional electronics by semiconductor defects. *Nano Today* **16**, 30-45 (2017).
24   Ye, M., Seo, H. & Galli, G. Spin coherence in two-dimensional materials. *npj Computational Materials* **5**, 44 (2019).
25   Chu, R.-L. *et al.* Spin-orbit-coupled quantum wires and Majorana fermions on zigzag edges of monolayer transition-metal dichalcogenides. *Physical Review B* **89**, 155317 (2014).
26   Edelberg, D. *et al.* Approaching the Intrinsic Limit in Transition Metal Diselenides via Point Defect Control. *Nano Letters* **19**, 4371 (2019).
27   Haastrup, S. *et al.* The Computational 2D Materials Database: high-throughput modeling and discovery of atomically thin crystals. *2D Materials* **5**, 042002 (2018).
28   Ataca, C., Sahin, H. & Ciraci, S. Stable, single-layer MX2 transition-metal oxides and dichalcogenides in a honeycomb-like structure. *Journal of Physical Chemistry C* **116**, 8983-8999 (2012).
29   Li, W.-F., Fang, C. & van Huis, M. A. Strong spin-orbit splitting and magnetism of point defect states in monolayer $WS_2$. *Physical Review B* **94**, 195425 (2016).
30   Wang, D. *et al.* Electronic and magnetic properties of MoS2 monolayers with antisite defects. *Journal of Physics and Chemistry of Solids* **131**, 119-124 (2019).
31   Wang, D. *et al.* Determination of Formation and Ionization Energies of Charged Defects in Two-Dimensional Materials. *Physical Review Letters* **114**, 196801 (2015).
32   Xia, S. *et al.* Evaluation of Charged Defect Energy in Two-Dimensional Semiconductors for Nanoelectronics: The WLZ Extrapolation Method. *Annalen der Physik* **532**, 1900318 (2020).
33   Dreyer, C. E., Alkauskas, A., Lyons, J. L., Janotti, A. & Van de Walle, C. G. First-principles calculations of point defects for quantum technologies. *Annual Review of Materials Research* **48**, 1-26 (2018).
34   Chernikov, A. *et al.* Exciton binding energy and nonhydrogenic Rydberg series in monolayer WS(2). *Physical Review Letters* **113**, 076802 (2014).
35   Choi, S., Jain, M. & Louie, S. G. Mechanism for optical initialization of spin in NV− center in diamond. *Physical Review B* **86**, 41202 (2012).
36   Maze, J. R. *et al.* Properties of nitrogen-vacancy centers in diamond: the group theoretic approach. *New Journal of Physics* **13**, 025025 (2011).
37   Alkauskas, A., Buckley, B. B., Awschalom, D. D. & Van De Walle, C. G. First-principles theory of the luminescence lineshape for the triplet transition in diamond NV centres. *New Journal of Physics* **16**, 073026 (2014).
38   Doherty, M. W., Manson, N. B., Delaney, P. & Hollenberg, L. C. L. The negatively charged nitrogen-vacancy centre in diamond: the electronic solution. *New Journal of Physics* **13**, 025019 (2011).
39   Bassett, L. C., Alkauskas, A., Exarhos, A. L. & Fu, K.-M. C. Quantum defects by design. *Nanophotonics* **8**, 1867-1888 (2019).
40   Ivády, V., Abrikosov, I. A. & Gali, A. First principles calculation of spin-related quantities for point defect qubit research. *npj Computational Materials* **4**, 76 (2018).

41  Klimeš, J., Bowler, D. R. & Michaelides, A. Chemical accuracy for the van der Waals density functional. *Journal of Physics: Condensed Matter* **22**, 022201 (2009).
42  Román-Pérez, G. & Soler, J. M. Efficient Implementation of a van der Waals Density Functional: Application to Double-Wall Carbon Nanotubes. *Physical Review Letters* **103**, 096102 (2009).
43  Ghasemi majd, Z., Amiri, P. & Taghizadeh, S. F. Ab-initio study of structural and electronic properties of $WS_2$/h-BN van der Waals heterostructure. *Surface Science* **672**, 13-18 (2018).
44  Kresse, G. & Furthmüller, J. Efficiency of ab-initio total energy calculations for metals and semiconductors using a plane-wave basis set. *Computational Materials Science* **6**, 15-50 (1996).
45  Kohn, W. & Sham, L. J. Self-consistent equations including exchange and correlation effects. *Physical Review* **140**, A1133 (1965).
46  Hohenberg, P. & Kohn, W. Inhomogeneous electron gas. *Physical Review* **136**, B864 (1964).
47  Blöchl, P. E. Projector augmented-wave method. *Physical Review B* **50**, 17953-17979 (1994).
48  Joubert, D. From ultrasoft pseudopotentials to the projector augmented-wave method. *Physical Review B* **59**, 1758-1775 (1999).
49  Heyd, J., Scuseria, G. E. & Ernzerhof, M. Hybrid functionals based on a screened Coulomb potential. *Journal of Chemical Physics* **118**, 8207-8215 (2003).
50  Krukau, A. V., Vydrov, O. A., Izmaylov, A. F. & Scuseria, G. E. Influence of the exchange screening parameter on the performance of screened hybrid functionals. *Journal of Chemical Physics* **125**, 224106 (2006).
51  Wu, Q. & Van Voorhis, T. Direct optimization method to study constrained systems within density-functional theory. *Physical Review A* **72**, 024502 (2005).
52  Kaduk, B., Kowalczyk, T. & Van Voorhis, T. Constrained density functional theory. *Chemical Reviews* **112**, 321-370 (2012).

# Supplementary Information

**Antisite defect qubits in monolayer transition metal dichalcogenides**

Jeng-Yuan Tsai[1], Jinbo Pan[1], Hsin Lin[2], Arun Bansil[3]*, Qimin Yan[1]*

[1]Department of Physics, Temple University, Philadelphia, PA 19122, USA

[2]Institute of Physics, Academia Sinica, Taipei, Taiwan

[3]Physics Department, Northeastern University, Boston, MA 02115, USA

*Correspondence and requests for materials should be addressed to Q. Y. (qiminyan@temple.edu) and A.B. (ar.bansil@northeastern.edu)

## 1. Pristine and defective 1H-TMDs

The ordering of $d_{x^2-y^2}, d_{xy}$, and $d_{z^2}$ levels, lattice constants, and data on the energetic of the triplet and singlet states of neutral anion antisites in six 1H-TMDs are presented in Table S1.

**Table S1** Geometric parameters in pristine $MX_2$ and defect systems with neutral anion antisites $M_X^0$. $d_{M-X}$ denotes the distance between a cation and its adjacent anions. $d_{X-X}$ is the distance between anions in upper and lower sublattices. In the defected systems, distances between the antisite cation (labeled A) and the three adjacent cations are labeled by $M_1$, $M_2$, and $M_3$. Distances between the antisite cation A and the anion in the bottom layer are labeled by $X_{below}$. The $d$-orbit-order refers to the ordering energy (increase) of in-gap defect levels. Defect levels are mainly composed of $d$ orbitals of antisite cations. Note, $M_{Te}^0$ in $MoTe_2$ and $WTe_2$ hosts a set of levels labeled as $d_{z^2}$, $(d_{xy}, d_{xz})$, and $(d_{yz}, d_{x^2-y^2})$, where $(d_{xy}, d_{xz})$ and $(d_{yz}, d_{x^2-y^2})$ denote hybridization between the two $d$ orbitals referenced.

| Pristine TMD | $WS_2$ | $WSe_2$ | $WTe_2$ | $MoS_2$ | $MoSe_2$ | $MoTe_2$ |
|---|---|---|---|---|---|---|
| a (Å) | 3.147 | 3.275 | 3.505 | 3.147 | 3.278 | 3.505 |
| $d_{M-X}$ (Å) | 2.391 | 2.513 | 2.701 | 2.385 | 2.511 | 2.700 |
| $d_{X-X}$ (Å) | 3.107 | 3.311 | 3.578 | 3.090 | 3.300 | 3.573 |
| Antisite $M_X^0$ | $W_S^0$ | $W_{Se}^0$ | $W_{Te}^0$ | $Mo_S^0$ | $Mo_{Se}^0$ | $Mo_{Te}^0$ |
| $d_{A-M_1}$ (Å) | 2.612 | 2.590 | 2.472 | 2.665 | 2.654 | 2.460 |
| $d_{A-M_2}$ (Å) | 2.612 | 2.590 | 2.672 | 2.665 | 2.654 | 2.769 |
| $d_{A-M_3}$ (Å) | 2.612 | 2.590 | 2.471 | 2.665 | 2.654 | 2.460 |
| $d_{A-X_{below}}$ (Å) | 3.488 | 3.562 | 2.935 | 3.544 | 3.631 | 2.961 |
| d orbital order | $d_{xy}, d_{x^2-y^2}, d_{z^2}$ | $d_{xy}, d_{x^2-y^2}, d_{z^2}$ | $d_{z^2}, (d_{xy}, d_{xz}), (d_{yz}, d_{x^2-y^2})$ | $d_{xy}, d_{x^2-y^2}, d_{z^2}$ | $d_{xy}, d_{x^2-y^2}, d_{z^2}$ | $d_{z^2}, (d_{xy}, d_{xz}), (d_{yz}, d_{x^2-y^2})$ |
| $\Delta E_{triplet-singlet}$ (eV) | -0.275 | -0.290 | -0.523 | -0.350 | -0.345 | -0.458 |

## 2. Transition levels and ionization energies of antisite defects $M_X^q$ in 1H-TMDs

**Anisotropic corrections for charged defects.** The anisotropic correction method for charged defect systems is based on the extrapolation of an asymptotic expression of ionization energy (IE):[1,2] $IE(S, Lz) = IE_0 + \frac{\alpha}{\sqrt{S}} + \frac{\beta}{S} L_z$, where $IE_0$ is size-independent ionization energy, S is the area of the 2D system, and $L_z$ is the vacuum thickness. Here, $\alpha$ is the Madelung constant for a point charge and mainly depends on geometry instead of defect type.[1,2] The supercells with areas of 5×5 and 6×6 unit cells and vacuum thicknesses of 20 and 25 Å, respectively, for each charged defect $M_X^q$ with charge state q = -1, 0, and 1 were employed. The anisotropic correction was carried out in two steps: (1) We extrapolated $IE(S, L_z)$ with respect to $L_z$. The y-intercept of the linear line represents $IE_0 + \frac{\alpha}{\sqrt{S}}$. Note that there are two linear lines for 5×5 and 6×6 supercells; followed by, (2) we extrapolated $IE_0 + \frac{\alpha}{\sqrt{S}}$ with respect to $\frac{1}{\sqrt{S}}$. Here the y-intercept is $IE_0$, the size-independent ionization energy. Note that ionization energy of the donor-state $\epsilon(+/0)$/acceptor-state $\epsilon(0/-)$ is defined as the energetic difference between the donor state/acceptor state and the CBM/VBM.

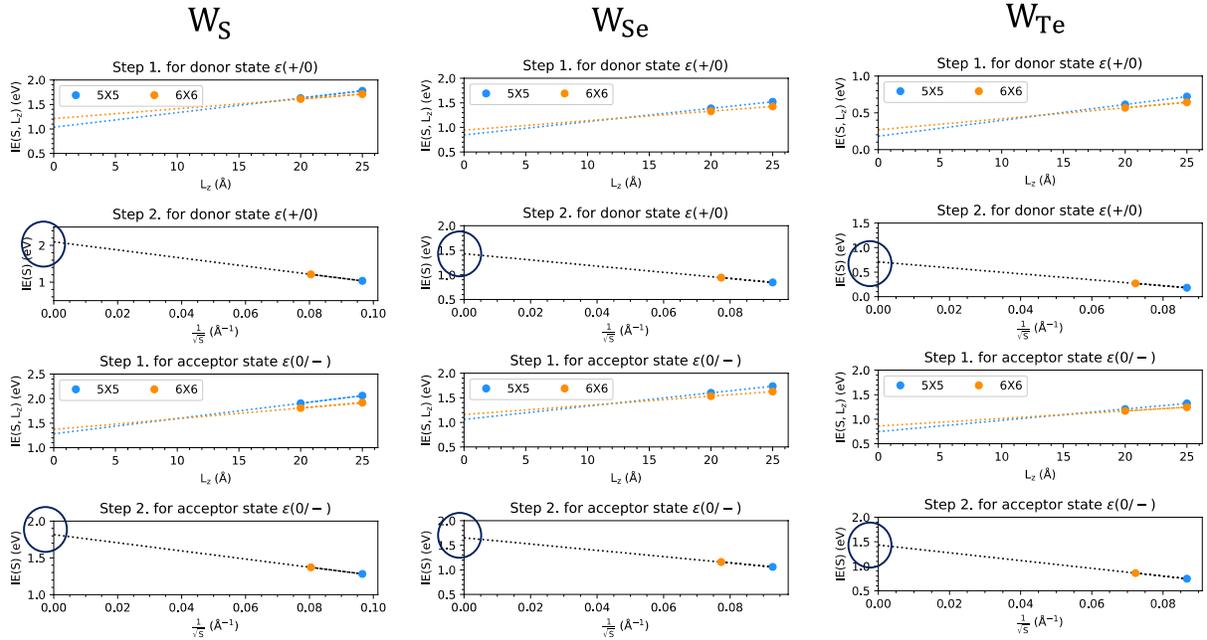

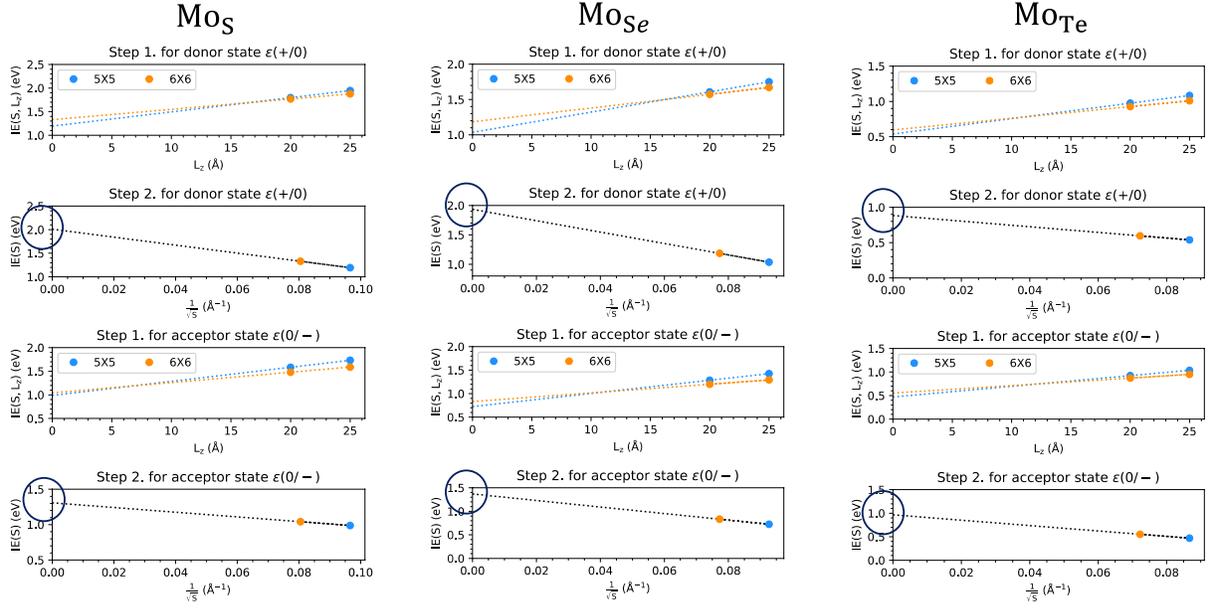

**Figure S1** Circles highlight the size-independent ionization energies ($IE_0$) of the donor $\epsilon(+/0)$ and acceptor states $\epsilon(0/-)$ for various antisites as follows. $W_S$: 2.10 eV, 1.82 eV; $W_{Se}$: 1.43 eV, 1.65 eV; $W_{Te}$: 0.71 eV, 1.43 eV; $Mo_S$: 2.01 eV, 1.31 eV; $Mo_{Se}$: 1.93 eV, 1.37 eV; and, $Mo_{Te}$: 0.88 eV, 0.97 eV.

## 3. Symmetry-allowed intersystem crossings in $W_S^0$

The single-particle spin-orbit operator $H_{so}$ with $C_{3v}$ symmetry can be defined as $H_{so} = \sum_k \lambda_{xy}(l_k^x s_k^x + l_k^y s_k^y) + \lambda_z l_k^z s_k^z$, where $l_k^i$ and $s_k^i$ denote the angular momentum and spin operators projected on to the i[th] component for the k[th] electron, respectively.[3] The nonaxial and axial strengths of spin-orbit coupling are represented by $\lambda_{xy}$ and $\lambda_z$, respectively. One can rewrite $H_{so}$ in terms of the raising and lowering operators of angular momenta in the form: $H_{so} = \sum_k \lambda_{xy}(l_k^+ s_k^- + l_k^- s_k^+) + \lambda_z l_k^z s_k^z$, where $l_k^\pm$ and $S_k^\pm$ are the raising and lowering operators for the angular momentum and spin operator, respectively. Nonaxial components include $l_\pm$ and $s_\pm$ that mix states with different single Slater determinants (i.e. mix $e^2$ and $a_1^1 e^1$) and different spin projections. On the other hand, the axial component contains $l_z$ and $s_z$ that is only able to mix states involving the same single Slater determinants and spin projections. We emphasize that $NV^-$ center hosts identical sublevel symmetries and electron/hole configurations. As a result, the nonaxial spin-orbit interaction is much weaker than the axial one in the $NV^-$ center.[4]

Intersystem crossing is allowed only if $<\Psi_i|H_{SO}|\Psi_f>$ is nonzero. Since $H_{so}$ is a scalar operator, which belongs to irrep $A_1$, its matrix element would be non-zero only if the irreps involved satisfy the condition: $\text{rep}(\Psi_i) \otimes \text{rep}(H_{so}) \otimes \text{rep}(\Psi_f) \supset A_1$. For the triplet ground state, the spatial wavefunction $\psi_{\text{spatial}}$ belongs to irrep $A_2$ and the three spin projectors $\{S_x, S_y\}$ and $S_z$ belong to the 2D irrep E and 1D irrep $A_2$, respectively. By decomposing the reducible representation $(E \oplus A_2) \otimes A_2$ into irreps of $C_{3V}$, we obtain the 2D degenerate sublevels E and the 1D sublevel $A_1$. The dimension of the sublevel space is three given by the product of the dimensions of $\psi_{\text{spatial}}$ (1D) and the spin projectors (3D). Accordingly, for the triplet excited state $^3$E, we have a six-dimensional sublevel space consisting of irreps $\{E_{12}, E_{xy}, A_1, A_2\}$. Here, the irreps $\{E_{12}, A_1, A_2\}$ involves spin components $\{S_x, S_y\}$ while the irrep $E_{xy}$ only involves $S_z$. The singlet states $^1$E and $^1A_1$ form sublevels labeled by E and $A_1$, which share a common spin-zero component ($S_0$). A nonzero matrix element also depends on the conditions that the axial (nonaxial) spin-orbit operators mix sublevels with identical (distinct) single Slater determinants and spin projectors. As a result, we identified three allowed intersystem crossings as noted in the main text.

## 4. Antisite qubit $W_S^0$ with local symmetry $C_h$

If a random symmetry-breaking structural perturbation is applied, $W_S^0$ antisite may experience an in-plane displacement from the center, which would lower the symmetry from $C_{3v}$ to $C_h$. Note that lowering $C_{3v}$ to $C_h$ reduces the total energy by only about 25 meV per unit cell. The results of $W_S^0$ with local symmetry $C_h$ are presented in this section.

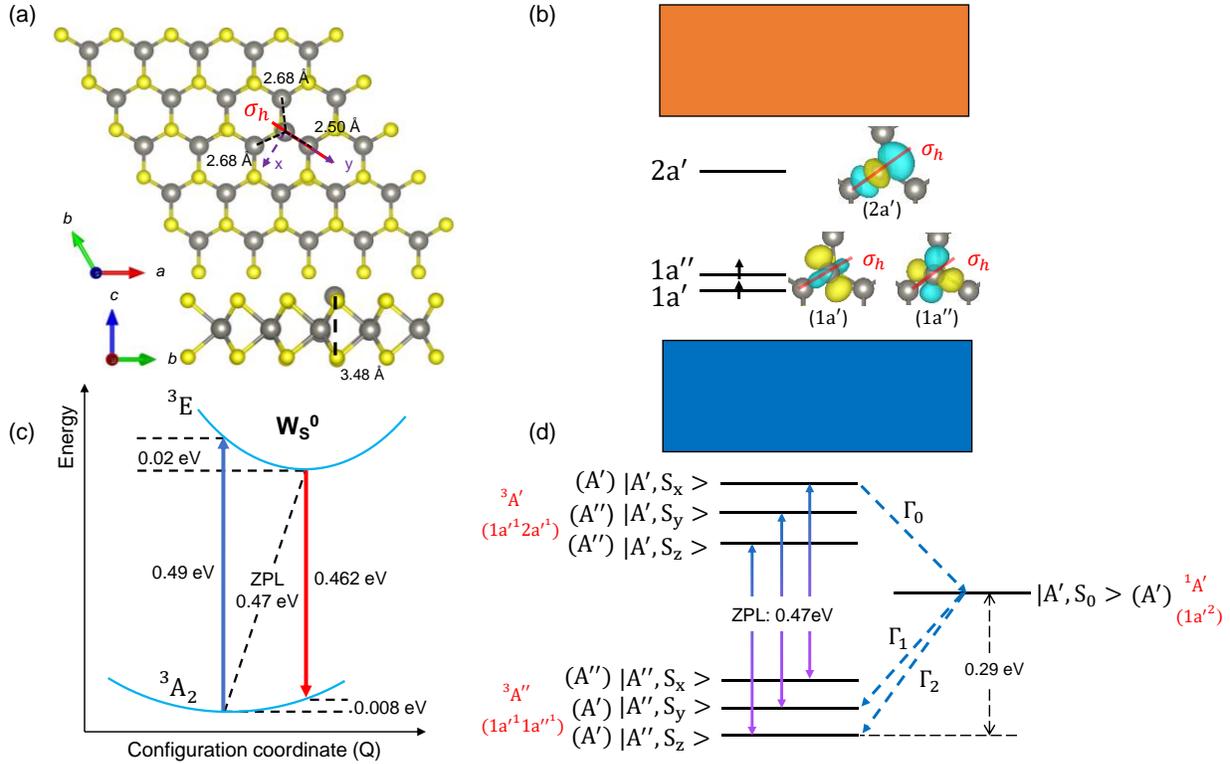

**Figure S2 (a)** The relaxed structure of $W_S^0$ with local symmetry $C_h$. *y*-direction is defined to lie along the mirror plane $\sigma_h$ (red line) and the *x*-direction is perpendicular to $\sigma_h$. **(b)** A schematic energy diagram of in-gap defect levels and the corresponding wavefunctions obtained via first-principles computations. Wavefunctions are labeled by irreps $a'$, $a''$, and $a'$ of $C_h$ from low to high energies. Despite hybridization with the neighboring *d* orbitals, the lowest unoccupied level is mainly contributed by $d_{z^2}$ orbital of the antisite. **(c)** Calculated configuration-coordinate diagram showing the zero-phonon-line (ZPL) and Frank-Condon relaxation energies of 0.47 eV and 8 meV, respectively. **(d)** Sublevels and the corresponding irreps for the triplet ground state $^3A'$, triplet excited state $^3A''$, and the singlet state $^1A'$. Symmetry-allowed intersystem-crossing decay paths $\Gamma_0, \Gamma_1$, and $\Gamma_2$ including effects of the spin-orbit coupling (SOC).

## Operation of the antisite defect $W_S^0$ with $C_h$ local symmetry

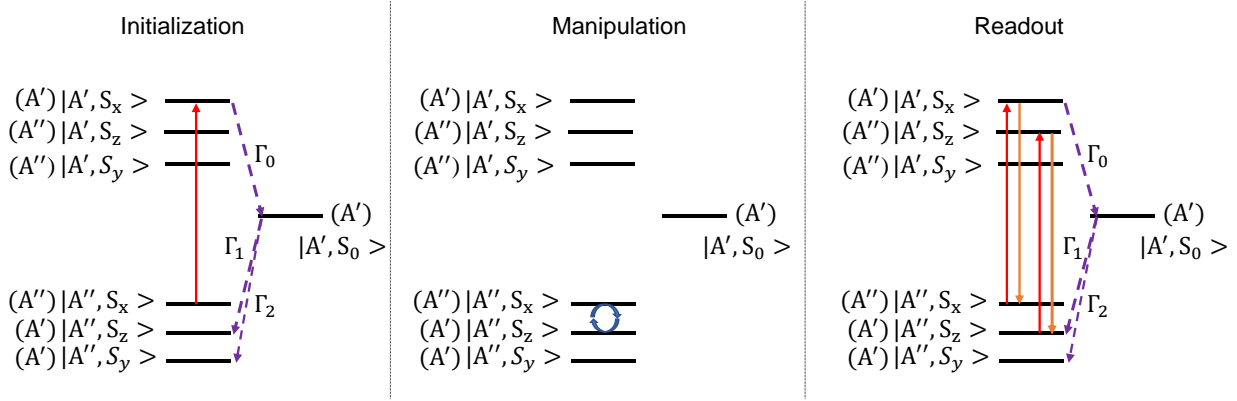

**Figure S3** Initialization (left panel) of the qubit involves adjusting and ordering of the positions of sublevels $S_x(m_s = 1)$, $S_z(m_s = 0)$, and $S_y(m_s = -1)$ by applying a strong external magnetic field. The increased energy gap between the sublevels $|A'', S_y >$ and $|A', S_0 >$ will suppress the intersystem crossing path $\Gamma_2$. Initialization of the qubit can be then achieved by pumping the antisite to the triplet excited state. Due to the existence of intersystem crossing paths $\Gamma_0$ and $\Gamma_1$ majority of the population will stay in the sublevel $|A'', S_z >$ of the triplet ground state. Manipulation of the qubit (middle panel) can be implemented via techniques similar to those used in $NV^-$ centers such as electron paramagnetic resonance (EPR). The readout process can also be implemented in analogy with $NV^-$ by pumping the antisite first and then detecting intensity differences in luminescence. Note that luminescence intensity from $|A', S_x > \rightarrow |A'', S_x >$ is dimmer than that for the $|A', S_z > \rightarrow |A'', S_z >$ transition due to the existence of the intersystem crossing path $\Gamma_0$.

**References**

1   Wang, D. *et al.* Determination of Formation and Ionization Energies of Charged Defects in Two-Dimensional Materials. *Physical Review Letters* **114**, 196801 (2015).
2   Xia, S. *et al.* Evaluation of Charged Defect Energy in Two-Dimensional Semiconductors for Nanoelectronics: The WLZ Extrapolation Method. *Annalen der Physik* **532**, 1900318 (2020).
3   Yu, P. Y. & Cardona, M. Fundamentals of Semiconductors. Physics and Materials Properties. *Springer* (2005).
4   Manson, N. B., Harrison, J. P. & Sellars, M. J. Nitrogen-vacancy center in diamond: Model of the electronic structure and associated dynamics. *Physical Review B* **74**, 104303 (2006).